\documentclass[%
 reprint,
 superscriptaddress,
 amsmath,amssymb,
 aps,
 pra,
]{revtex4-2}
\usepackage{hyperref}
\usepackage{amsmath}
\usepackage{xcolor}
\usepackage{graphicx}
\usepackage{siunitx}
\usepackage{bm}

\usepackage{datetime}

\date{\today}

\makeatletter
\def\maketitle{
\@author@finish
\title@column\titleblock@produce
\suppressfloats[t]}
\makeatother

\begin{document}

\title{Long optical coherence times and coherent rare earth-magnon coupling in a rare earth doped anti-ferromagnet}
\author{Masaya Hiraishi}
\author{Zachary H. Roberts}
\author{Gavin G. G. King}
\author{Luke S. Trainor}
\affiliation{%
  Department of Physics, University of Otago, Dunedin, New Zealand}
\affiliation{Dodd-Walls Centre for Photonic and Quantum Technologies, Dunedin, New Zealand}
\author{Jevon J. Longdell}
\email{jevon.longdell@otago.ac.nz}
\affiliation{%
  Department of Physics, University of Otago, Dunedin, New Zealand}
\affiliation{Dodd-Walls Centre for Photonic and Quantum Technologies, Dunedin, New Zealand}
\affiliation{
Quantum Technologies and Dark Matter Labs, Department of Physics, University
of Western Australia, 6009 Crawley, Australia}

\begin{abstract}
  
Rare-earth ions are characterised by transitions with very narrow linewidths even in solid state crystals. Exceedingly long coherence times have been shown on both spin and optical transitions of rare-earth-ion doped crystals. 
A key factor, and generally the limitation, for such coherence times, is the effects of electronic and nuclear spins in the host crystal. 
Despite the attractive prospect, a low-strain, spin-free host crystal for rare-earth-ion dopants has not yet been demonstrated. 
The dopants experience the lowest strain when they substitute for another rare earth (including yttrium). 
However every stable isotope of the trivalent rare earth ions has either an electron spin, an nuclear spin, or both. 
The long optical coherence times reported here with erbium dopants in antiferromagnetically ordered gadolinium vanandate suggest an alternative method to achieve the quiet magnetic environment needed for long coherence times: use a magnetic host fully concentrated in electron spins and operate at temperatures low enough for these spins to be ordered. 
We also observe avoided crossings in the optical spectra, caused by strong coupling between the erbium ions and gadolinium magnons in the host crystal. This suggests the exciting prospect of microwave to optical quantum transduction using the rare-earth ions in these materials mediated by magnons of the host spins.

\end{abstract}

\maketitle

Rare-earth ions doped in solid state crystals exhibit microwave \cite{Zhong_2015,rancic_coherence_2018,ortu_simultaneous_2018} and optical absorption spectra \cite{Bottger_2009,equall_ultraslow_1994} with very narrow linewidths and correspondingly long coherence times. This is because their 4$f$-to-4$f$ transitions are electrostatically shielded by the outer closed shells. At low temperatures the shielding is effective enough that the dominant dephasing mechanism for both optical and spin transitions is often magnetic fluctuations from spins in the host.

Much effort has been made to find rare-earth host crystals with as few electronic and nuclear spins as possible in order to reduce the magnetic fluctuations. At the same time it is desirable that the dopants substitute for a similar ion, both in terms of size and charge state, to reduce strain and the inhomogeneous linewidths in the crystal. For those reasons hosts where the rare-earth ion substitutes for another rare earth (including yttrium) lead to the narrowest inhomogeneous linewidths. 

Arguably the best low strain, low spin, host for rare-earth ions is currently yttrium orthosilicate (Y$_2$SiO$_5$). The host ion is Y$^{3+}$, which has no net electron spin, but it does have nonzero nuclear spin $I = 1/2$ for its only stable isotope ($^{89}$Y). Yttrium spins dominate the dephasing with oxygen and silicon only having nuclear spins for minor isotopes.
The longest coherence times among the rare-earth ions in solids have been achieved in host Y$_2$SiO$_5$ crystals on both spin \cite{Zhong_2015} and optical transitions  \cite{Bottger_2009,equall_ultraslow_1994}.

In addition to the suitable host-material selection, the effects of magnetic fluctuations can be further suppressed by the zero first-order Zeeman (ZEFOZ) technique, \cite{fraval_method_2004,wolfowicz_atomic_2013,mohammady_bismuth_2010,Zhong_2015}. A ZEFOZ point represents a particular applied magnetic field vector where a transition frequency is first-order insensitive to magnetic field fluctuations. By operating at such points dramatic extension of coherence times have been demonstrated \cite{Zhong_2015}.

Even with such a nearly spin-free host material and the ZEFOZ technique, the remaining spins still play a dominant role in a decoherence process. Spin-free, low strain hosts for rare earths have not been achieved, because an unfortunate coincidence means that no trivalent rare-earth ion has zero nuclear spin \emph{and} zero electronic spin. Every second trivalent rare-earth ion along the periodic table is a non-Kramers ion, having an even number of electrons and thus the possibility for frozen out electronic spins in low symmetry hosts. However all these non-Kramers ions have an odd atomic number which means that, the stable isotopes all have an odd atomic mass and therefore non-zero nuclear spin. The lack of even atomic mass non-Kramers ions is because of the low stability of ``odd-odd'' nuclei: those with odd numbers of both protons and neutrons.

This work was inspired by the impressive properties shown by crystals where the rare-earth ion of interest is no longer a dopant but present in stoichiometric quantities as part of the host crystal. 
One example of the potential of these `fully concentrated' rare-earth materials is the very long optical coherence time $T_2$ of Eu$^{3+}$ ions in a  {EuCl$_3 \cdot 6$D$_2$O} crystal \cite{Ahlefeldt_2013_optical} .

Europium is a non-Kramers ion and in ${\textrm{EuCl}_3 \cdot 6\textrm{H}_2\textrm{O}}$ has no electron spin.
Stoichiometric crystals containing \emph{Kramers} ions have their own unique properties; cooling them down to temperatures of $\sim\qty{1}{K}$ allows the electronic spins to order, resulting in a quiet magnetic environment despite all of the spins.
Quite recently, narrow inhomogeneous linewidths on optical transitions of antiferromagnetic erbium lithium fluoride (ErLiF$_4$) were reported \cite{Berrington_2023}, but there has still been little study of their optical coherence.

On microwave transitions, ultrastrong coupling between a microwave cavity resonator and a collective excitation of electronic spin waves, a magnon, was achieved with antiferromagnetic gadolinium orthovanadate (GdVO$_4$) \cite{Everts_2020}. 
The ultrastrong coupling was owing to the number of spins involved: the coupling strength is proportional to the square root of the number spins.
The linewidth of the magnon was also narrow at \qty{35}{MHz}, despite the sample's cuboidal shape.

Microwave to optical conversion for quantum states is a compelling prospect for enabling interconnection and scaling of superconducting qubit computers. A number of aproaches are being pursued \cite{Awschalom.2021,Lambert.2020,Wang.2022,Higginbotham.2018,Jiang.2020,Zhong.2020,Fan.2018,Rueda.2016,Tu.2022,Vogt.2019,Hisatomi.2016,Zhu.2020,Everts.2019}. 

This magnon rare-earth coupling provides a promising avenue for microwave--optical conversion. Recently microwave-optical conversion with rare earths achieved close to 1\% efficiency \cite{xie_scalable_2024} but the efficiency and bandwidth are limited by weak oscillator strengths for the microwave spin transitions.
If we can instead have a microwave cavity that is strongly coupled to a magnon and have that magnon strongly coupled to the rare-earth ion, we eliminate this weak link \cite{tharn23,tharn24}.

Here, we present an optical spectroscopic study of erbium (Er$^{3+}$) ions in an antiferromagnetic GdVO$_4$ crystal (Er:GdVO$_4$). We show the optical absorption spectra for the $^4I_{15/2}(Z_1)\rightarrow {}^4I_{13/2}(Y_1,Y_2)$ transitions and show that these can be explained by a model where the erbium dopants are affected by a crystal field from the host crystal, the static mean field of the gadolinium spins as well as coherent coupling to the gadolinium magnons. Strong coupling between erbium ions and magnon resonances are seen, with a fitted exchange coupling strength of \qty{-11.3}{\giga\hertz}.

\section{\texorpdfstring{E\lowercase{r}:G\lowercase{d}VO\textsubscript{4}}{Er:GdVO4}}
\label{sec:ergvo}

\begin{figure*}
    \centering  \includegraphics[width=1\linewidth]{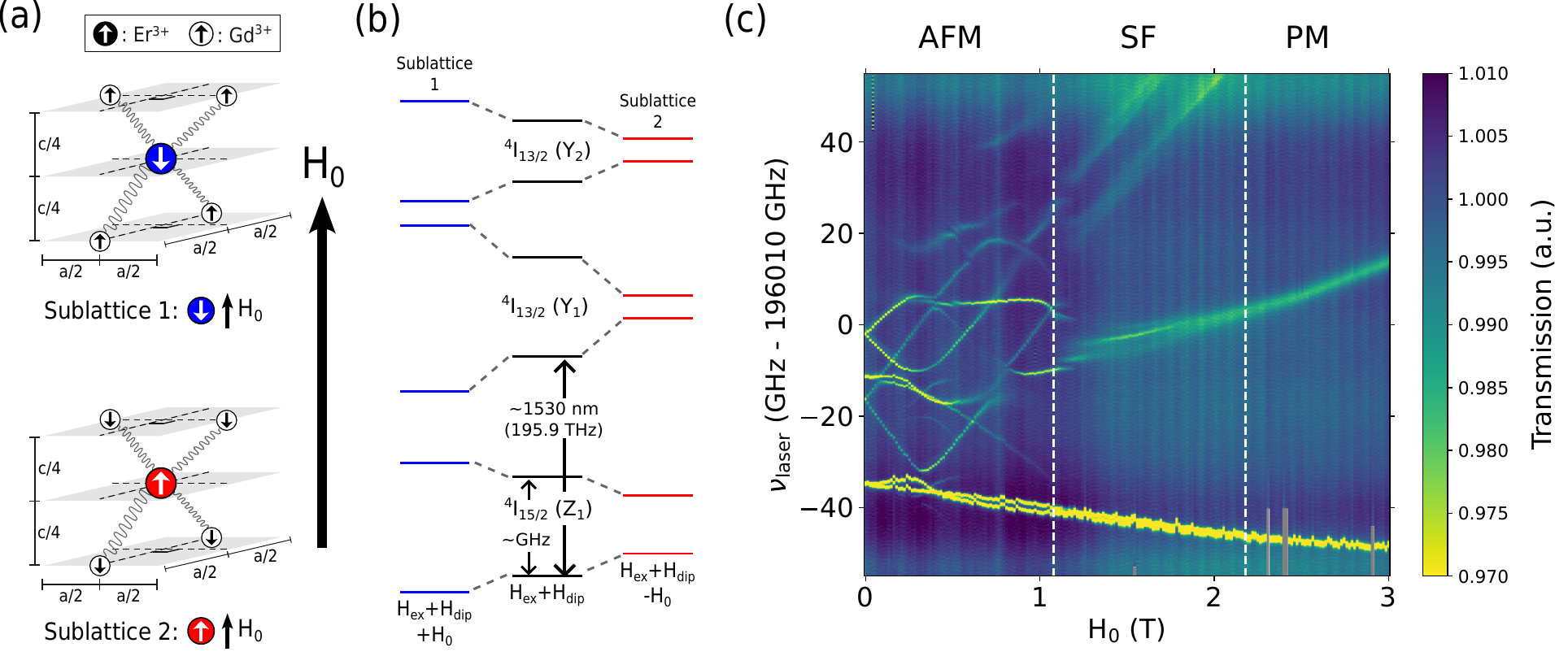}
    \caption{
    (a) The local environment of the erbium dopants.
    The erbium ions order antiferromagnetically to their four nearest-neighbour gadolinium ions in a $D_{2d}$ point group.
    The ordered erbium spin orientation defines our two sublattices.
    (b) Erbium energy levels of interest for this study.
    The black levels in the centre are the $Z_1$, $Y_1$ and $Y_2$ Kramers doublets, which are split by the mean exchange ($\textrm{H}_{\textrm{ex}}$) and magnetic dipole ($\textrm{H}_{\textrm{dip}}$) fields of the gadolinium. The $Y_1$ and $Y_2$ are close to each other and are mixed.
    An external magnetic field ($\textrm{H}_{0}$) is either parallel or anti-parallel to the gadolinium mean field, which causes the energy levels of the two sublattices (1, blue) and (2, red) to shift differently.
    (c) Optical transmission spectrum with light polarised halfway between $\sigma$ and $\pi$ polarisations, showing excitation from the lowest level of the $Z_1$  doublet to the doublets $Y_1$ and $Y_2$. Three different magnetic phases of GdVO$_4$ are labelled at the top; antiferromagnetic (AFM), spin-flop (SF), and paramagnetic (PM) phases. White dashed lines represent the magnetic phase transition fields of GdVO$_4$ from the AFM to the SF states (\qty{1.08}{\tesla}) and the SF to the PM states (\qty{2.18}{\tesla})  \cite{Cashion.1970}. Grey areas shown at the lower frequencies are areas where the optical transmission was not recorded properly. To show the weaker absorption lines more clearly, the data has been cut off at the limits shown in the colour bar. The strongest absorption measured was for $\sigma$-polarised light on the lowest transition; there we observed a peak absorption of \qty{26}{\percent}.
    }
    \label{fig: energy level and crystal structure}
\end{figure*}

The material used in this study is erbium-ion doped gadolinium vanadate (Er:GdVO$_4$), where the guest and host rare-earth ions are Er$^{3+}$ and Gd$^{3+}$, respectively. Our crystal was purchased as undoped GdVO$_4$ but was found to have impurities of other rare earths including erbium.

GdVO$_4$ is tetragonal with the Gd$^{3+}$ ion occupying site with $D_{2d}$ symmetry \cite{milligan_crystal_1952}. The crystal becomes antiferromagnetically ordered at the N\'eel temperature $T_N= 2.495$\,K \cite{Cashion.1970}. The magnetic structure is a simple two sublattice antiferromagnet with the easy axis along the $c$ axis, where each Gd$^{3+}$ ion is anti-parallel to all the four nearest neighbor Gd$^{3+}$ ions. Some Gd$^{3+}$ ions are substituted by Er$^{3+}$ ions as shown in Fig.\,\ref{fig: energy level and crystal structure}(a).

Figure \ref{fig: energy level and crystal structure}(b) shows a schematic of the energy-levels of Er$^{3+}$ ions in antiferromagnetic GdVO$_4$ relevant to our study. The optical transitions are from the lowest state of the ground doublet (labeled as $^4I_{15/2}(Z_1)$) to the optically excited states associated with the $^4I_{13/2}(Y_1)$ and $^4I_{13/2}(Y_2)$ doublets. The Er$^{3+}$ ions are expected to magnetically interact with neighboring Gd$^{3+}$ ions via an effective magnetic field produced by exchange and magnetic dipole-dipole interactions, which breaks the time-reversal symmetry (Kramers' degeneracy). 
When in the ordered state, an external magnetic field along the $c$ axis causes the Er$^{3+}$ ions to experience two different effective magnetic fields: in `sublattice 1' the external field is anti-parallel to the Er$^{3+}$ spin, while in `sublattice 2' the external field is parallel, as shown in Fig.~\ref{fig: energy level and crystal structure}(a).

\section{Optical Absorption Spectra}
\label{sec:optical:absorption}
Optical spectra with a linear-polarised input at \qty{45}{\degree} between $\pi$ (electric field $\textbf{E} \parallel$ the $c$ axis) and $\sigma$ ($\textbf{E} \perp c$) polarisations are shown in Fig.\,\ref{fig: energy level and crystal structure}(c). The applied magnetic field direction is along the $c$ axis, where a small misalignment of the field direction from the $c$ axis is likely to be at most $\theta \sim \qty{10}{\degree}$ as discussed in the Supplementary Information.
Polarisation dependence is seen on the observed transitions, indicating selection rules. The absorptions at the bottom [near \qty{-40}{\giga\hertz} in Fig.\,\ref{fig: energy level and crystal structure}(c)], corresponding to the transitions of the two sublattices from their lowest $Z_1$ to lowest $Y_1$ levels (Fig.\,\ref{fig: energy level and crystal structure}(a)), are mainly $\sigma$ polarised. This is consistent with Er:YVO$_4$, which has magnetic dipole transitions with the same polarisation \cite{Xie_2021, Li_2020}. The other transitions are mainly $\pi$ polarised.

We estimate that our Er$^{3+}$ concentration is $\sim\qty{1}{ppm}$ by comparing the optical absorption coefficient of this transition to Er:YVO$_4$, accounting for the ratio of the inhomogeneous linewidths.

The number of observed absorption lines can be explained by the magnetic phase of GdVO$_4$.  
At zero applied field, we see four absorption lines. Even with no applied field, there is an effective magnetic field at the Er$^{3+}$ ions, causing Zeeman splitting. The four lines correspond to transitions from the lowest level of the Zeeman $Z_1$ doublet to the two levels in each of the $Y_1$ and $Y_2$ doublets. Applying an external magnetic field along the $c$ axis removes the degeneracy between the two sublattices in the antiferromagnet, so that eight transitions become visible, four each from the ions in sublattices 1 and 2.

At about \qty{1.1}{\tesla} there is a phase transition between the antiferromagnetic phase and a spin-flop phase \cite{Page_1977, Everts_2020}.
In the spin-flop phase, the two sublattices point in opposite directions when projected onto the plane perpendicular with the $c$ axis, but are canted up an equal amount towards the $c$ axis \cite{Everts_2020}. For this phase we expected both sublattices to have the same transitions frequencies, but observed a small difference. We attribute this to misalignment of the external magnetic field from the $c$ axis.
Another transition, between the spin-flop and paramagnetic phases, is known to occur around \qty{2.2}{\tesla} \cite{Cashion.1970}. In the paramagnetic phase, the Er$^{3+}$ spins at the two sublattices are identical and are expected to have the same transition frequencies even with the misalignment, as was observed. 

The narrowest inhomogeneous linewidths observed were approximately \qty{386\pm 9}{\mega\hertz}, comparable to the \qty{163\pm14}{\mega\hertz} seen for the narrowest transition in the best samples of Er:YVO$_4$ \cite{Xie_2021}. 

In the antiferromagnetic phase a rich structure is apparent; the closeness of the $Y_1$ and $Y_2$ doublets means that they are mixed by the internal fields from the gadolinium host.
We numerically simulate the relevant spectra with two different models: first with a standard crystal field model, starting with parameters from Ref.~\cite{Bertini_2004}, with single isolated erbium ions interacting with an electric crystal field, as well as effective magnetic dipolar and exchange fields; and then as a system coupled to the host Gd$^{3+}$ ions. Details of the models and calculations are in Sec.~S5 of the Supplementary Information.

\begin{figure*}
    \centering
    \includegraphics[width=\linewidth]{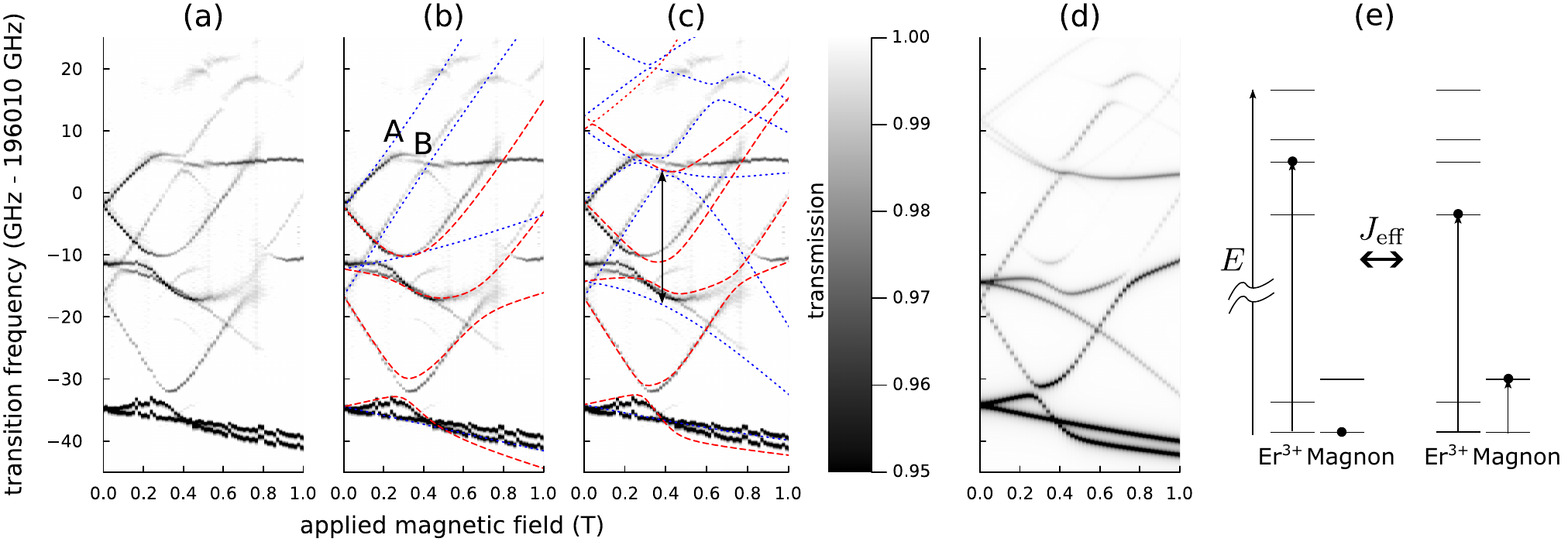}
    \caption{
    Measured absorption spectra compared with two models. 
    (a) The measured spectra for clarity.
    (b) The experimental data overlaid with the predictions from erbium under the influence of the lattice crystal field, static exchange fields and magnetic dipolar fields produced by the gadolinium, as well as the applied magnetic field.
    Blue dotted lines (red dashed lines) are the expected transition frequencies for sublattice 1 (2).
    A number of avoided crossings due to coupling between the erbium and the magnons are not predicted by this model.
    Prominently, at points `A' and `B' the transition frequencies are expected to linearly increase, but couple to something else. 
    (c) Extending the model to include coupling to the gadolinium magnons predicts a lot of structure not predicted by the simpler model.
    The avoided crossing marked with an arrow arises due to coupling between the erbium and magnons. Its fitted effective exchange strength is $J_\mathrm{eff}=\qty{-11.3}{GHz}$, and the avoided crossing is seen to be a splitting of $|2J_\mathrm{eff}|$.
    The coupling strength is sufficiently strong to reverse the magnetic-field sensitivity of the third transition frequency near zero field, which is observed in both experiment and theory. 
    (d) Calculated magnetic-dipole transition strengths from the magnon-coupled crystal field model, averaging $\sigma$ and $\pi$ polarisations. The fourth transition at about \qty{-2}{GHz} at zero applied field is not seen in these calculations as it is magnetic-dipole forbidden like in Er:YVO\textsubscript{4} \cite{Xie_2021}. For this graph we assume all transitions have the same linewidth though this will not be true especially at high frequencies.
    (e) Illustrates the two excited states of the combined system which are coupled to leading to the avoided crossing marked with an arrow in (c). The state where the erbium is excited to its second excited-state level couples strongly to the combined state with the erbium in its first excited-state level and a single magnon excitation.
    }
    \label{fig: simulated spectra}
\end{figure*}

Figure \ref{fig: simulated spectra} shows the measured spectra and the models used to explain it.
At low temperatures and with applied magnetic fields along the $c$ axis that are small compared to the $\approx \qty{1}{T}$ required for the spin flop phase, the magnetisation of the gadolinium stays constant. This simplifies modelling the erbium absorption spectrum because the effect of the static field of the gadolinium can be described by two numbers: the strength of the exchange field and the internal magnetic field. As shown in Fig.~\ref{fig: simulated spectra}(b) a crystal field model for the erbium with these two effective fields as well as the external magnetic field explain many features of the observed spectra. 
There are, however, a number of points where the mean-field model cannot account for the features observed. At points marked `A', `B' in Fig.~\ref{fig: simulated spectra}(b), the transition frequencies of sublattice 1 would be expected to linearly increase, but instead they show avoided crossings with some other energy level.
The avoided crossings can be explained as an interaction between the Er$^{3+}$ ions and the magnons. The model that includes these interactions also includes these avoided crossings, as shown in Fig.~\ref{fig: simulated spectra}(c) and explained in Sec.~S5 of the Supplementary Information.

\section{Optical phase memory time measurements}
\label{sec:coherence:optical:time}
\begin{figure}
    \includegraphics[width=\linewidth]{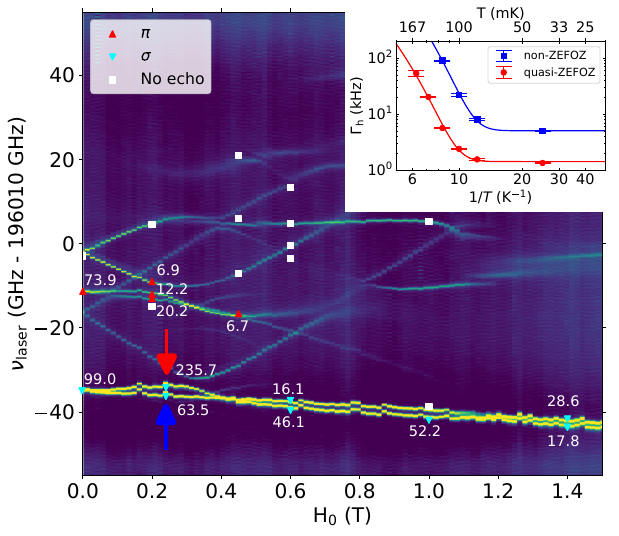}
    \caption{\label{fig: T2 temp dependence2} Measured photon echo phase memory times. Symbols denote points of measurement; squares show points where no echoes were observed, while triangles show measured phase memory times in microseconds for the polarisations shown. The background is the same optical spectrum as Fig.~\ref{fig: energy level and crystal structure}. The inset shows the temperature dependence of the two longest transitions indicated by arrows in the main plot.}
\end{figure}

Coherent properties of the Er$^{3+}$ optical transitions were investigated through two-pulse photon echo measurements. The observed echo decays are nonexponential, which is consistent with $^{167}$Er:YVO$_4$ \cite{Li_2020}. We fitted the results with a Mims decay function for such nonexponential echo amplitude decays: $\exp[-(2t_{12}/T_M)^{x}]$ where $t_{12}$ is the pulse delay between the first and second pulses, and $x$ is the stretch factor. The fitted phase memory times $T_M$ are shown in Fig.~\ref{fig: T2 temp dependence2} at the points in the spectrum where they occur.

Note that there are spots shown where we attempted to observe echoes but no echoes were observed. These are mostly $Z_1 \rightarrow Y_2$ transitions with high magnetic-field sensitivity. These transitions thus seem to have $T_M$ times less than the \qty{5}{\micro\second} the experiment could resolve. The $1\sigma$ uncertainties for the fitted values for $T_M$ were at most \qty{11}{\percent}.

The longest $T_M$ was measured on the lowest$\rightarrow$lowest $Z_{1} \rightarrow Y_{1}$ transition of sublattice 2 at \qty{0.24}{\tesla}. Near this point, the magnetic field dependence of the transition frequency seems smaller than the other points, indicating that the point is insensitive to magnetic fluctuation from the neighbouring spins, which is close to a ZEFOZ point.
The observed $T_M = \qty{235.7\pm 5.9}{\micro\second}$ is comparable with $T_M = \qty{336}{\micro\second}$ achieved with isotopically purified $^{167}$Er$^{3+}$ doped YVO$_4$ measured at \qty{1.0}{\tesla} and below \qty{1}{\kelvin} \cite{Li_2020}.

With this longest $T_M$ transition, the temperature dependence of $T_M$ was investigated. 
For comparison, we measured this temperature dependence at each of the two sublattices. 
We do not have a perfect ZEFOZ transition, because the crystal was not aligned perfectly along the applied magnetic field. Nonetheless, the transition does have a reduced sensitivity to the magnetic field variations, and we will call it a `quasi-ZEFOZ' transition, to compare with the non-ZEFOZ transition.

The inset of Fig.~\ref{fig: T2 temp dependence2} shows the homogeneous linewidths $\Gamma_{{\rm h}}$ that we defined as $\Gamma_{{\rm h}} \equiv (\pi T_M)^{-1}$.
The homogeneous linewidths on both transitions are insensitive to temperature at low temperatures then show rapid broadening with increasing the temperature above $\sim$\qty{100}{\milli\kelvin}.

Here, we hypothesise a model that explains the observed temperature dependence of the homogeneous linewidths, i.e., a model for a homogeneous linewidth of an optically-active rare-earth ion doped into a magnetically ordered crystal.

We decide to model the homogeneous linewidth as a function of temperature with
\begin{equation}\label{eqn:linewidths}
    \Gamma_{{\rm h}}= \Gamma_{0} + \Gamma_{\Delta} \exp\left(\frac{-h\Delta}{k_B T}\right),
\end{equation}
where $k_B$ and $h$ are the Boltzmann and Planck constants; $\Gamma_{0}$ is the residual linewidth reached in our experiments in the limit of low temperature; $\Gamma_{\Delta}$ describes the strength of broadening caused by the surrounding electron spins; and $\Delta$ corresponds to the energy required to excite these spins.

With this model, the temperature-dependent dephasing is explained by spin-spin interaction between the single Er$^{3+}$ spin and the Gd$^{3+}$ magnon. The magnon is thermally excited, which is incorporated as the exponential term, $\exp(-h\Delta / k_B T)$. $\Gamma_{\Delta}$ then represents how much the electron spin-spin interaction affects the line broadening.

The residual linewidth $\Gamma_0$ we attribute to magnetic fluctuations that aren't frozen out even at our lowest temperatures, most likely nuclear spins in the host. This is supported by the fact that the ratio $\Gamma_\Delta/\Gamma_0$ is similar for both transitions, as shown in Table~\ref{tab:fits}.

This equation was fitted to the data in the inset of Fig.~\ref{fig: T2 temp dependence2} to make the curves; the parameters for the best fit to Eq.~\ref{eqn:linewidths} are shown in Table~\ref{tab:fits}.
No echoes were observed above \qty{160}{\milli\kelvin} at the quasi-ZEFOZ point and \qty{140}{\milli\kelvin} at the non-ZEFOZ.
\begin{table}
    \caption{\label{tab:fits}
        Parameters to Eq.~\ref{eqn:linewidths} shown in the inset of Fig.~\ref{fig: T2 temp dependence2}.
    }
    \begin{tabular}{c c c c}
        \hline\hline
                                & Quasi-ZEFOZ   & Non-ZEFOZ     &   \\
        \hline
        $\Gamma_{0}$            & $1.4\pm0.1$   & $5.1\pm0.3$   & kHz\\
        $\Gamma_{\Delta}$       & $38\pm22$     & $160\pm74$    & MHz\\
        $\Delta$                & $22\pm1.6$    & $19\pm1.1$    & GHz\\
        $\Gamma_\Delta/\Gamma_0$& $(26\pm15)$   & $(31\pm14)$   & $\times10^{3}$\\
        \hline\hline
    \end{tabular}
\end{table}

While large uncertainty is obtained specifically on $\Gamma_{\Delta}$, which is likely due to the small number of the data points, the estimated two $\Delta$ are close to each other, and its value is also close to the lower Gd$^{3+}$ magnon mode of 
$\simeq\qty{26}{\giga\hertz}$ at \qty{0.24}{\tesla} \cite{Abraham_1992}. 

\section{Discussion}
\label{sec:discussion}

We explained the rich structure in our optical spectra without hyperfine structure. One isotope of erbium, \textsuperscript{167}Er, does have a nuclear spin and should be present at \qty{23}{\percent} natural abundance.
We see no clear evidence of \textsuperscript{167}Er\textsuperscript{3+} in our optical spectra, which we attribute to the strong bias field from the gadolinium and the symmetry of the experiment. The strong field along the $c$ axis will cause the \textsuperscript{167}Er\textsuperscript{3+} energy levels to be close to eigenvectors of the nuclear spin $I_z$ operator, reducing the oscillator strength for a transition between hyperfine levels.
Nevertheless, we observed long term hole-burning in our sample with a three-pulse echo technique at the quasi-ZEFOZ point.
Repeated application of two laser pulses of length \qty{1.5}{\micro\second} spaced by \qty{5}{\micro\second} created a frequency grating with free spectral range \qty{200}{\kilo\hertz} in the absorption profile of the ions. A third pulse \qty{20}{\second} later was observed to create an echo, showing that some of the \textsuperscript{167}Er had decayed into other nuclear spin states creating persistent spectral holes. The \qty{20}{\second} time period shows that the hyperfine $T_1$ is at least a similar order of magnitude as the \qty{15}{\second} observed in YVO\textsubscript{4} \cite{Li_2020}. This long lifetime demonstrates the potential for \textsuperscript{167}Er:GdVO\textsubscript{4} to be used for quantum memories.

Rare-earth ions in solids are under investigation as a route to microwave to optical quantum signal conversion. \cite{Bartholomew.2020,  OBrien.2014,
Williamson.2014,  Welinski.2019,King.2021, fernandez-gonzalvo_cavity-enhanced_2019}. 
In this approach the bottleneck for the conversion process is usually the weak coupling between the rare-earth spins and microwave photons \cite{xie_scalable_2024}.
Moving to crystals where the rare-earth ion is part of the crystal rather than being a dopant has been proposed \cite{Everts.2019} and (ultra)strong coupling between a cavity mode and a rare-earth magnon mode demonstrated \cite{Everts_2020}. One issue with these ``fully concentrated'' crystals is that the coupling between the rare-earth ions and the optical cavity is extreme, with large dispersion making phase matching difficult. Various schemes for coupling magnons to rare earths have been suggested, in transition metal based magnets \cite{tharn23} and with a rare-earth doped crystal close to a magnetically ordered system \cite{tharn24}. Our multi-gigahertz coupling between the Er$^{3+}$ dopants and the Gd$^{3+}$ magnons make this material and ones like it an exciting prospect for magnon-enhanced rare-earth upconversion to be realized. 

The coherence of rare-earth-ion dopants is usually greatly improved with the application of a bias magnetic field which grows the ``frozen core'' of host spins \cite{szabo} surrounding the dopant.
One advantage of an antiferromagnetic rare-earth system like ours over a non-magnetic system such as YVO$_4$ is the possibility for this bias field to come from within the crystal introducing the prospect of long optical coherence times and microwave-to-optical conversion with zero applied magnetic field. 

An advantage over transition metal magnets is the typically weak superexchange interaction seen in rare-earth antiferromagnets. In particular, the strength of the ordering interaction is large enough to order the spins at millikelvin temperatures but, in contrast to transition metal magnets \cite{Rezende_2019}, the ordering interaction is weak enough that the magnons are in the same range as superconducting qubits with zero or small applied magnetic fields.

\section{Acknowledgements}

The authors would like to thank R.~L.~Ahlefeldt, J.~G.~Bartholomew, and S.~Rogge for discussions.
We would also like to thank M.~J.~Sellars and R.~L.~Ahlefeldt for the use of some equipment.

This work was supported by the NZ Ministry of Business, Innovation and Employment (MBIE) Catalyst Strategic New Zealand–German Aerospace Centre Joint  Research Programme Contract No. UOOX2112.

This work was supported by Quantum Technologies Aotearoa, a research programme of Te Whai Ao -- the Dodd Walls Centre, funded by the New Zealand Ministry of Business Innovation and Employment through International Science Partnerships, contract number UOO2347.

\renewcommand{\selectlanguage}[1]{}

\pagebreak

\title{Supplementary Information for \texorpdfstring{\\}{ } Long optical coherence times and coherent rare earth-magnon coupling in a rare earth doped anti-ferromagnet}

\maketitle

\setcounter{section}{0}
\setcounter{equation}{0}
\setcounter{figure}{0}
\setcounter{table}{0}
\makeatletter
\renewcommand{\thesection}{S\arabic{section}}
\renewcommand{\theequation}{S\arabic{equation}}
\renewcommand{\thefigure}{S\arabic{figure}}
\renewcommand{\thetable}{S\arabic{table}}
\renewcommand{\bibnumfmt}[1]{[S#1]}
\renewcommand{\citenumfont}[1]{S#1}

\section{Sample and Mounting Overview}
\label{sec:sample_overview}

\begin{figure}
\includegraphics[width=\columnwidth]{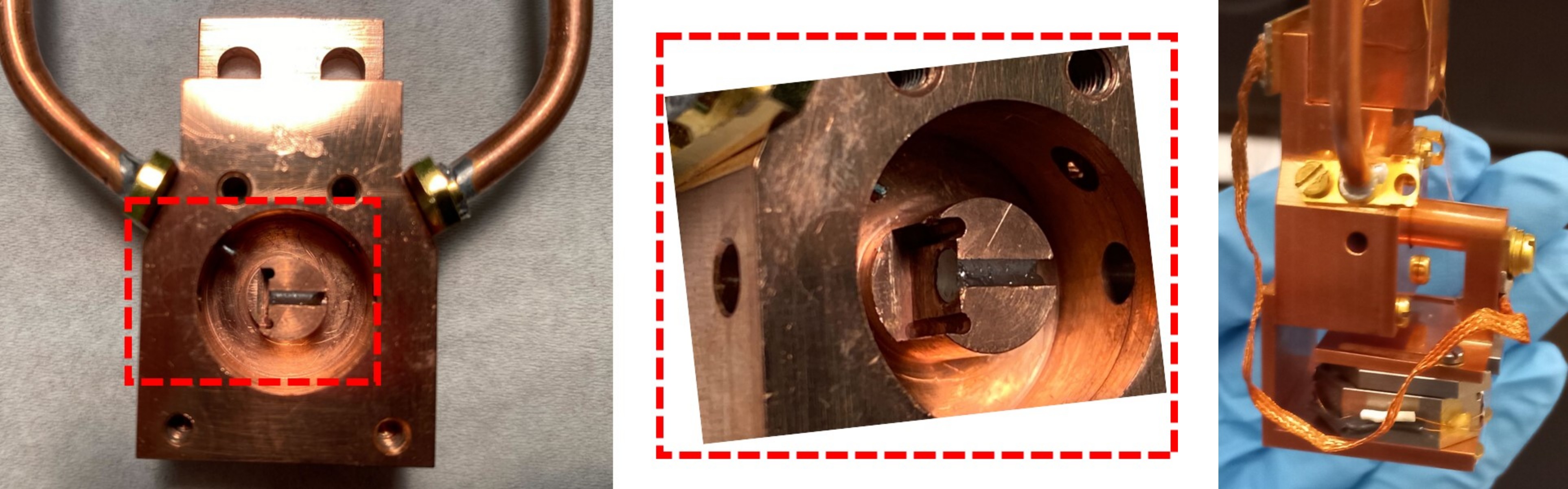}
\caption{\label{Fig: Sample and cavity} Left: Sample configuration in a frequency-tunable loop-gap microwave cavity resonator. The red dashed square represents a magnified photograph shown in the middle. Right: Assembled cavity attached to the cold finger of the DR. The copper tuning plunger is seen on the right.
}
\end{figure}

The sample was a cylinder of gadolinium vanadate of 2.4\,mm diameter and length 4.2\,mm. It was purchased as undoped, but contains  approximately 1\,ppm of erbium. The circular ends are normal to the $a$ axis, with the other $a$ axis and the $c$ axis in the plane of the ends. The sample was mounted in a tunable microwave loop-gap resonator with optical access through the circular ends of the sample, as shown in Figure~\ref{Fig: Sample and cavity}. The microwave cavity was made of oxygen-free copper for thermal and electrical conductivity at cryogenic temperatures. To tune the microwave cavity there was a plunger that was moved by a micro-actuator (Attocube ANPx101) to adjust the size of the gap in the loop gap resonator. The sample was mounted as a loose fit in the central hole, with thermal grease (Apiezon ``N'') for thermal contact.
The microwave cavity enabled us to observed the gadolinium magnon mode, which helped confirm the sample orientation.

The entire resonator was mounted on a cold finger attached to the coldest stage of a dilution refrigerator (BlueFors LD-250) so that it was in the centre of a 3\,T superconducting magnet. The magnet has optical access orthogonal to the magnetic field, with several layers of windows all anti-reflection coated for approximately 1540\,nm.

The orientation of the $c$ axis was determined by measuring the birefringence of the sample with crossed linear polarisers. The maximum misalignment between the $c$ axis and the magnetic field was $10^\circ$.  
The crystal orientation with applied microwave magnetic field $\perp c$ axis is important to efficiently couple to the low-frequency magnon's bulk magnetization.

\section{Optical spectroscopy}
\label{sec:supplement:opticalspectroscopy}
\label{sec: Optical transmission spectroscopy} 

\begin{figure}
    \centering
    \includegraphics[width=0.9\linewidth]{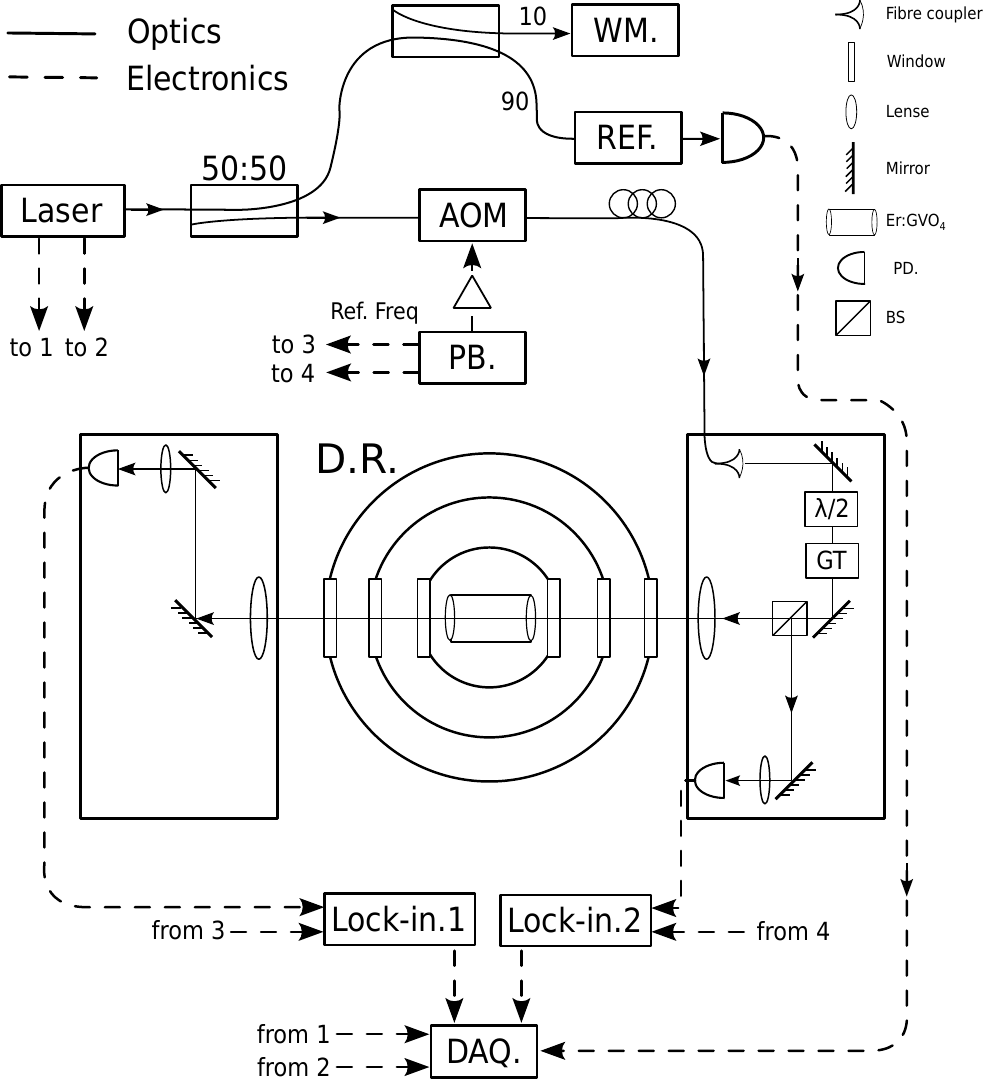}
    \caption{Schematic of the experimental setup of optical transmission. Lines of optics and electronics are represented by solid and dashed lines, respectively. \textbf{WM.}\,is a wavemeter; \textbf{REF.}\,is a reference-cavity system with a single-loop optical fibre splitter; \textbf{AOM} is an acousto-optic modulator; \textbf{PD.} is a photodetector; \textbf{PB.}\,is PulseBlaster, a radio-frequency generator; \textbf{Lock-in} is a lock-in amplifier; \textbf{$\lambda/2$} a half-wave plate; \textbf{GT} is a Glan-Taylor polariser; \textbf{D.R.}\,is a dilution refrigerator; and \textbf{DAQ.} is a data acquisition device.}
    \label{fig: setup optical transmission}
\end{figure}
Optical transmission spectra were measured with the configuration shown in Figure~\ref{fig: setup optical transmission}. The magnetic field was stepped, a single optical sweep was performed, and then the next magnetic field was stepped.

The optical source was a tunable diode laser module (Pure Photonics PPCL560) with a home-built microprocessor control allowing a triangular sweep of $\pm60$\,GHz from the set centre frequency.  The centre frequency was measured with a zero-span sweep, then once sweeping a fibre reference cavity comprising a fibre coupler coupled to itself was used to linearise the frequency scan.

The polarisation of the light was controlled by rotating the roughly polarised light out of the diode module, then passing it through a linear polariser.  

The power on the sample was approximately \qty{5.7}{\micro\watt}, a compromise between signal-to-noise ratio and sample heating.  The optical power varied as the laser swept, so the transmitted light was normalised to a reference beam picked off before the fridge.  Lock-in detection of both the sample and reference beams was used to further improve the signal to noise ratio.  The optical power was controlled by a $80$\,MHz AOM modulated with a square wave at 4.273\,kHz. The detectors were free-space InGaAs photodiodes (Hamamatsu G12180-010A).

The temperature of the mixing chamber plate was measured as 40\,mK throughout measurements of the optical spectra shown in Fig.~\ref{fig: energy level and crystal structure}(c).

\section{Microwave Spectroscopy}
\label{sec:supplement:microwavespectroscopy}
\label{sec:microwave_transmission}

\begin{figure}
\includegraphics[width=0.5\columnwidth]{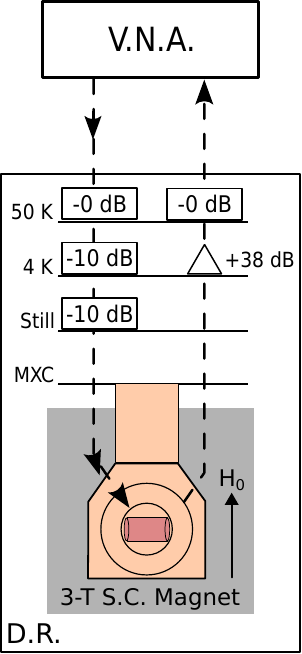}
\caption{\label{Fig: microwave setup} Experimental setup of the microwave transmission measurements. The cold finger and the microwave cavity resonator are magnified for clarity. Dashed lines represent coaxial cables with arrows indicating the microwave directions. Gray shaded area represents, labelled as 3-T S.C.\,Magnet, the \qty{3}{T} superconducting magnet. The arrow indicates the static field $\mathbf{H}_0$ direction. \textbf{V.N.A.}\,is a vector network analyser and \textbf{D.R.}\,is a dilution refrigerator.
}
\end{figure}
The sample is mounted in a microwave cavity that was designed to tune over approximately an octave from 6\,GHz to 12\,GHz with $Q\approx2000$.  These loop-gap resonators are known to have a filling factor of around \qty{80}{\percent} \cite{Fernandez-Gonzalvo.2015}, which allows stronger coupling between the microwave input field and the spins in both the gadolinium host and the erbium dopant. 

Transmission was measured with a vector network analyser (Siglent SNA5032A), although for these measurements the phase is not signifcant, as shown in Figure~\ref{Fig: microwave setup}. Several stages of microwave attenuation and amplification are used to reduce the amount of thermal noise reaching the microwave resonator. 

\begin{figure*}
\includegraphics[width=2\columnwidth]{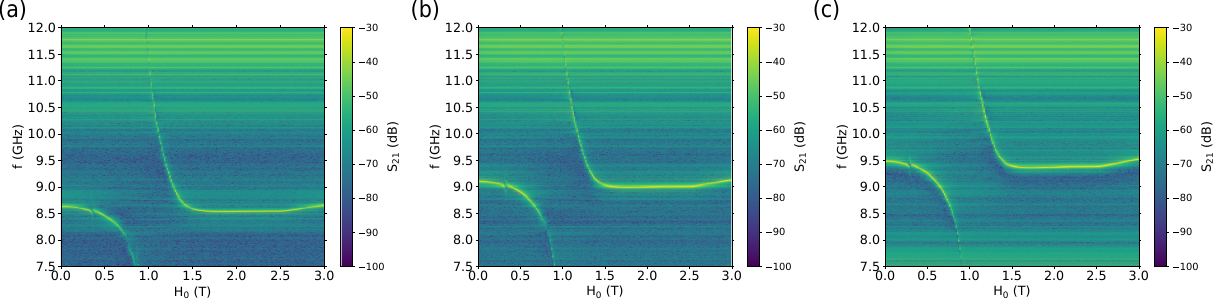}
\caption{\label{Fig: S21} Microwave transmission spectra for varying magnetic fields. The cavity was tuned to (a) 9.52\,GHz, (b) 9.11\,GHz, and (c) 8.64\,GHz.
}
\end{figure*}

In the past this resonator has been seen to tune over the entire design range \cite{hiraishi2023spectroscopy}, but for these measurements the plunger or actuator was stuck, limiting the tuning to between 8.64\,GHz and 9.52\,GHz. Transmission as a function of microwave frequency and magnetic field is shown in Figure~\ref{Fig: S21} for three plunger positions, giving three roughly evenly spaced centre frequencies. The most prominent feature is a strong avoided crossing at around \qty{1}{T} applied field, which is consistent with previously recorded measurements of gadolinium vanadate in a microwave resonator \cite{Everts_2020_S}. 

There is also a much smaller avoided crossing visible at around 0.3\,T. This was not seen in the measurements of pure gadolinium vanadate \cite{Everts_2020_S}, suggesting that it results from something other than the gadolinuim ions. It is not predicted by our numerical simulations of erbium doped into gadolinium vanadate, suggesting that it is not the erbium ions we know are present from optical spectroscopy. Because we know separation of rare earth ions is difficult, and that the sample was contaminated with erbium, we assume that this secondary avoided crossing is from some other rare earth impurity in the crystal.

\section{Two-pulse photon echo measurement}
\label{sec:photon_echo}
\begin{figure}
    \centering
    \includegraphics[width=0.9\linewidth]{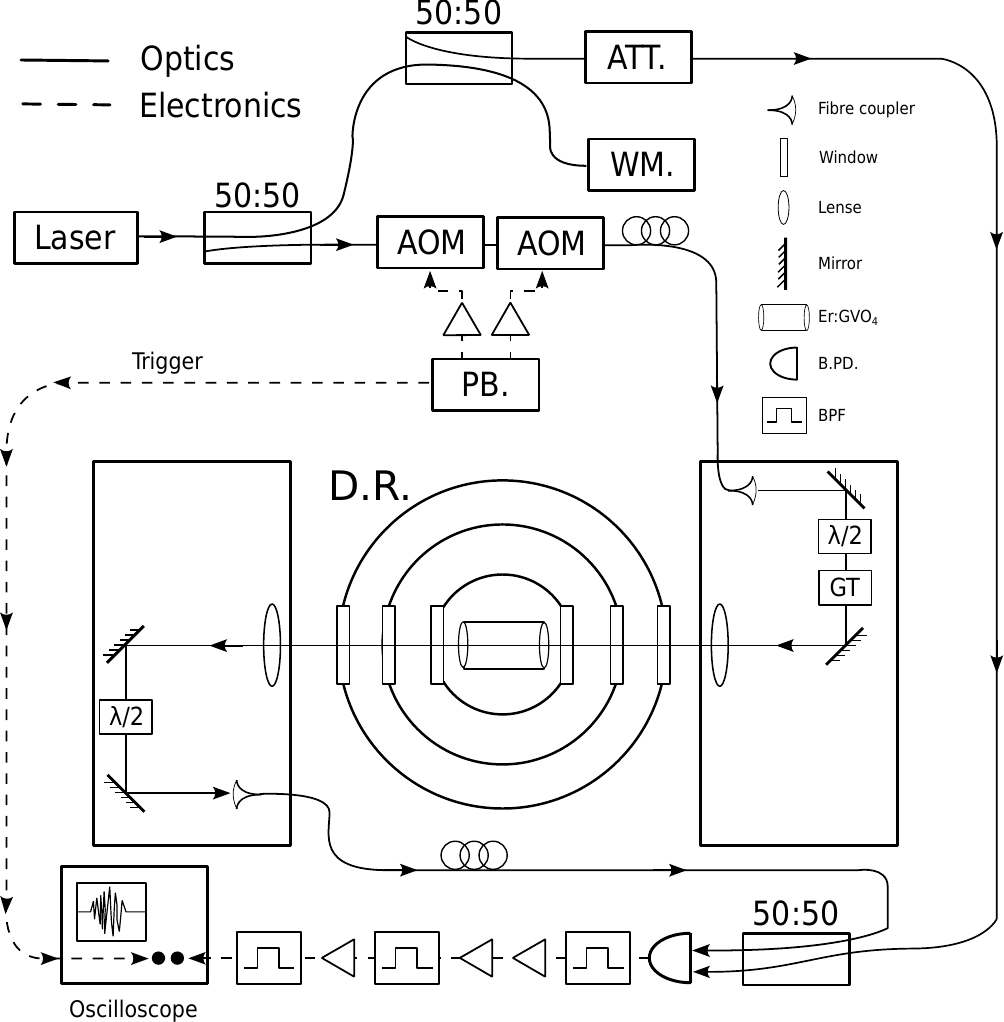}
    \caption{
    Experimental setup for two-pulse and three-pulse photon echo measurements. Lines of optics and electronics are represented by solid and dashed lines, respectively. Table optic components were put on two optical tables set just before and after the dilution refrigerator. An oscilloscope was used to monitor detected echo amplitude by a balanced photodetector (\textbf{B.PD.}). \textbf{AOM} denotes an acousto optic modulator; \textbf{ATT.}~is optical attenuator; \textbf{WM.}~is wavemeter; \textbf{PB.}~is PulseBlaster, a radio-frequency generator;\textbf{BPF} is a band-pass filter; \textbf{D.R.} is dilution refrigerator; \bm{$\lambda/2$} is half-wave plate; and \textbf{GT} is Glan-Taylor polariser.
    }
    \label{fig: setup photon echo}
\end{figure}

The two-pulse photon echo measurements were performed at various optical-transition field-frequency points as shown in Figure~\ref{fig: T2 temp dependence2}. 
They were measured with an optical setup shown in Figure \ref{fig: setup photon echo}.
For high-sensitivity detection, we employed optical heterodyne detection in a similar way as in Ref.~\cite{Rakonjac_2020}. The local oscillator is taken by preparing another optical line that bypasses two acousto-optic modulators (AOMs) and the crystal. Two free-space AOMs are put in series before the sample in order to create optical pulses.
They have up- and down-frequency shifts. We adjusted them to obtain the optical frequency of the transmitted light shifted by \qty{+10.8}{\mega\hertz}. Echoes couple to a fibre coupler and are mixed with the local oscillator by a 50:50 optical fibre splitter. The two outputs of the splitter are connected to a balanced photodetector with its bandwidth \qty{200}{\mega\hertz}. The 10.8-MHz beat signal with the local oscillator and the echo signal was amplified with several 10.8-MHz band-pass filters. The amplified and filtered beat signals were monitored by an oscilloscope for data collection. Since we monitor the beat signal between the echo and the local oscillator, what we actually measured was echo amplitude rather than intensity \cite{Rakonjac_2020}.

We first searched echoes at various observed optical transition points by setting a short pulse delay between the two pulses. At every optical transition point where the echo was observed, the $\pi/2$- and $\pi$-pulse widths were determined such that the echo amplitude was maximised. The input optical power was \qty{1}{\milli\watt} measured at continuous wave. With such optimised pulse-condition, the echoes were measured with various pulse-delay $t_{12}$ to investigate decay of the echo amplitude. We recorded echoes ten times at each $t_{12}$ because the used laser had frequency fluctuations that reduced the observed echo signals. Using the best echo amplitudes at each measurement setting is known to accurately reproduce the true coherence decay \cite{strickland_laser_2000}.

\begin{figure}
    \centering
    \includegraphics[width=\linewidth]{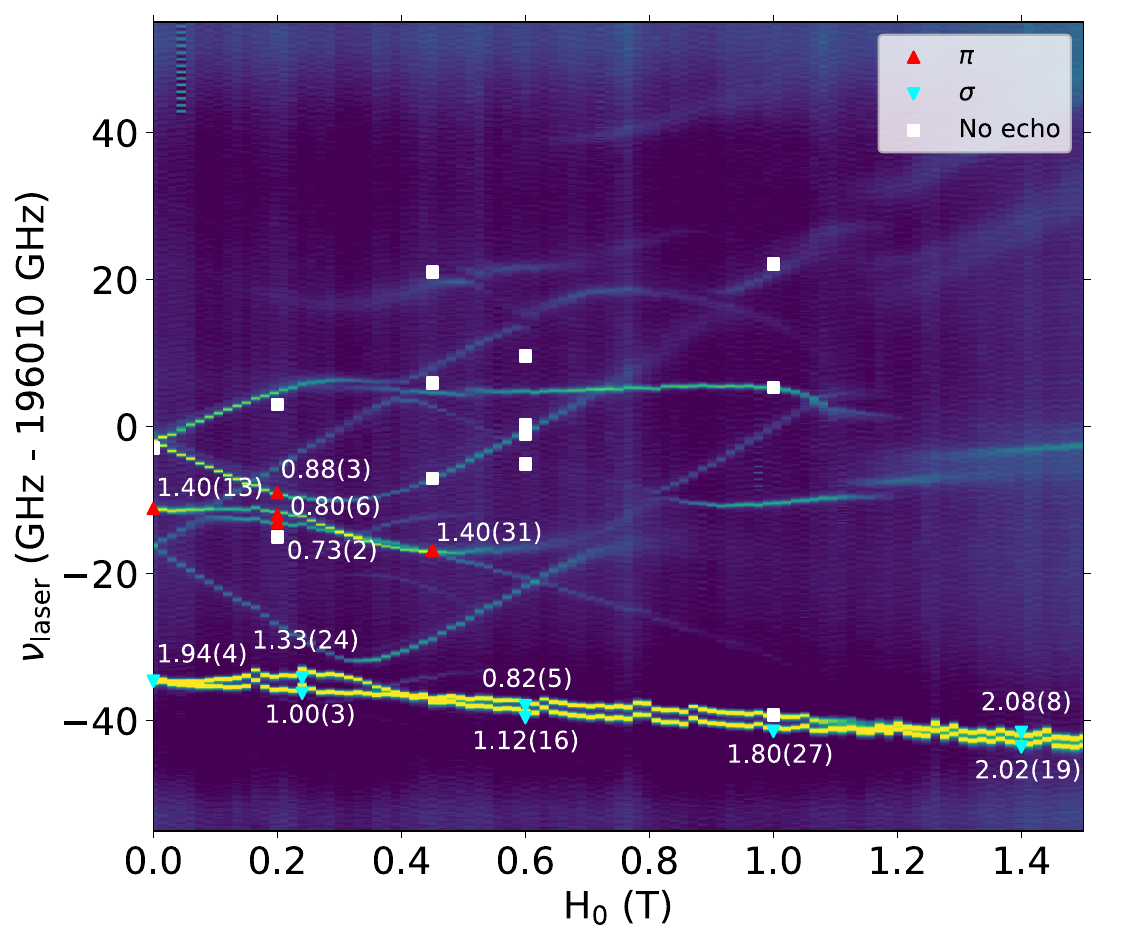}
    \caption{
    Stretch factor $x$ obtained by the fitting $\exp[-(2t_{12}/T_M)^{x}]$ to the measured two-pulse photon echo decays. The corresponding phase memory time $T_2$ are shown in Figure~\ref{fig: T2 temp dependence2} 
    of the main manuscript.
    }
    \label{fig: photon echo Mims factor map}
\end{figure}

The observed echo decays with $t_{12}$ exhibited different features from one measurement field-frequency point to another. They are exponential and nonexponential as observed in many rare earth doped materials including $^{167}$Er:YVO$_4$ \cite{Li_2020_S}. The nonexponential dependence has been known due to spectral diffusion because of local magnetic fluctuation in the crystal \cite{mims1968phase, Bottger_2006}. We fitted the observed echo decays with the decay function proposed by Mims \textit{et al.} to include spectral diffusion effects known as $\exp[-(2t_{12}/T_M)^{x}]$, where $x$ is the stretch factor that indicates dynamics of the spectral diffusion and $T_M$ is a phase memory time \cite{mims1968phase}. Figure \ref{fig: photon echo Mims factor map} shows the stretch factor $x$ determined at each measured echo, and the corresponding $T_M$ are shown in Figure~3 of the main manuscript.

\section{Crystal field model}
\label{sec:crystal-field-model}

The starting point for our crystal field model a refit to previous data from Reference \cite{Bertini_2004_S}, except we corrected one energy level that we believe to be a typo $\qty{6527}{\per\centi\meter}\rightarrow\qty{6537}{\per\centi\meter}$. We added our measured frequencies of the $Y_1$ and $Y_2$ transitions as a function of applied field, which we will refer to as our ``high-resolution data''. 

We had to discriminate between transitions of sublattices 1 and 2, which we did by keeping track of transitions that appear to avoid each other with varying applied field. The avoidance is either due to coherent coupling or misaglignment of the applied field.
The model requires extra terms due to the magnetic ordering. In the first instance we assumed no coupling to magnons. Our Hamiltonian was
\begin{equation}
    H_\mathrm{Er} = H_\mathrm{FI} + H_\mathrm{CF} + H_\mathrm{Z} + H_\text{Er MF} + H_{{}^4 I_{13/2}\text{ corr.}},
\end{equation}
where $H_\mathrm{FI}$ is the free ion Hamiltonian for erbium, and $H_\mathrm{CF}$ is the crystal field Hamiltonian. $H_\mathrm{Z}$ is the Zeeman interaction Hamiltonian given by
\begin{equation}
    H_\mathrm{Z} = \frac{\mu_B}{h} \bm{B} \cdot (\bm{L} + 2\bm{S}),
\end{equation}
where $\bm{L}$ and $\bm{S}$ are vectors of the erbium's angular momentum and spin operators. The applied magnetic field was taken to be along the $z$ axis with a small misalignment along $x$, $\bm{B} = B(\cos\theta\bm{\hat{z}} + \sin\theta\bm{\hat{x}})$

To account for an interaction with the mean field from the ordered neighbouring gadolinium spins, we incorporated two extra terms due to mean-field coupling via magnetic dipole--dipole ($B_\mathrm{dip}$) and exchange ($B_\mathrm{ex}$) interactions:
\begin{equation}
    H_\text{Er MF} = \frac{\mu_B}{h} [ B_\mathrm{ex}S_z + B_\mathrm{dip} (L_z+2S_z) ].
\end{equation}
Because GdVO$_4$ is an easy-axis uniaxial antiferromagnet and we are applying a field approximately along the easy axis, we have assumed the the mean field terms are static and along the easy axis ($z$ or $c$).

In our crystal-field fit, we apply a much harsher penalty to the high resolution fit than to the low resolution fit (zero-field data from Ref.~\cite{Bertini_2004_S}).
Additionally, as we are fitting field dependence, for the high-resolution data 
we must fit the energy difference between the ground state and the eigenvalue. To still maintain accuracy for the other levels, we add a projective correction Hamiltonian that offsets the ${}^4I_{13/2}$ manifold
\begin{equation}
    H_{{}^4 I_{13/2}\text{ corr.}} = E_\mathrm{corr.}\sum_{i,j\in {}^4 I_{13/2}} \lvert \psi_i \rangle \langle \psi_j \rvert,
\end{equation}
where $\lvert\psi_i\rangle$ are the eigenvectors at zero applied field. In practice we calculate this Hamiltonian (except for the coefficient $E_\mathrm{corr.}$) before optimization as it will remain constant as long as we do not vary the parameters far enough to mix the $J$ states.
The standard method uses a global energy offset $E_0$, which brings the ground state close to, but not exactly to zero energy. To maintain compatibility with this method the correction energy for our field-dependent data was a pragmatic and efficient choice.

We fit the low-field (applied field less than \qty{0.5}{T}) transition frequencies we measured for the $Z_1\rightarrow Y_1,Y_2$ transitions as well as the measured frequencies from Ref.~\cite{Bertini_2004_S}. We excluded points where the transitions had deviated in frequency due to magnon coupling. Our crystal field model only contained four new parameters: $B_\mathrm{ex}$, $B_\mathrm{dip}$, $\theta<\qty{10}{\degree}$, and $|E_\mathrm{corr.}| < \qty{10}{\per\centi\meter}$.

To get the model that we ultimately used to explain the observed spectrum in Er:GdVO$_4$, we added the gadolinium magnon and coupling terms to the above crystal field model, but importantly only one free parameter. The new Hamiltonian was:
\begin{equation}
    H = H_\mathrm{Er} + H_\mathrm{Gd} + H_\text{Er--Gd}.
\end{equation}
The gadolinium magnon is modelled as a spin-deviation of the nearest-neighbour gadolinium sublattice. We begin with the Zeeman Hamiltonian of the gadolinium
\begin{equation}
    H_\mathrm{Gd} = \frac{\mu_B}{h} g_\mathrm{Gd} [ \bm{B}\cdot\bm{S}_{\mathrm{Gd}} + B_\mathrm{0,Gd}S_{\mathrm{Gd},z} ],
\end{equation}
where $\bm{S}_\mathrm{Gd}$ is a vector of spin operators for the gadolinium sublattice; $B_\mathrm{0,Gd} \approx -1.2$\,T, the magnitude of which is the mean gadolinium exchange field; and $g_\mathrm{Gd}= 2$,  which is measured from the magnon field dependence and is expected considering Gd\textsuperscript{3+} has no orbital angular momentum in the ground state. 

We then substitute the spin-deviation creation and annihilation operators, $a^\dag$ and $a$, for this sublattice \cite{Rezende_2019_S}:
\begin{align}
    S_{\mathrm{Gd},x} &= \frac{\sqrt{N_\mathrm{Gd}}}{2}\left[ \left(I - \frac{a^\dag a}{N_\mathrm{Gd}}\right)^{1/2} a + a^\dag \left(I - \frac{a^\dag a}{N_\mathrm{Gd}}\right)^{1/2} \right]\nonumber \\
    S_{\mathrm{Gd},y} &= \frac{i\sqrt{N_\mathrm{Gd}}}{2}\left[  \left(I - \frac{a^\dag a}{N_\mathrm{Gd}}\right)^{1/2} a - a^\dag \left(I - \frac{a^\dag a}{N_\mathrm{Gd}}\right)^{1/2} \right]\nonumber \\
    S_{\mathrm{Gd},z} &= \frac{N_\mathrm{Gd}}{2} I - a^\dag a 
    \rightarrow - a^\dag a.
\end{align}
The final substitution is made because the term $(N_\mathrm{Gd}/2)I$ is a constant offset, and will be accounted for by the mean exchange field when we introduce coupling to the gadolinium. Here $N_\mathrm{Gd}$ is the total number of magnons possible in the system, i.e. the number of gadolinium ions in that sublattice. When we evaluate the eigenvalues, we must use a truncated Hilbert space for the magnons. The operators on the gadolinium subspace are matrices of size $(N_\mathrm{Gd}+1)\times(N_\mathrm{Gd}+1)$.

The interaction between the erbium and gadolinium magnon is taken to be an effective Heisenberg exchange interaction of exchange constant $J_\mathrm{eff}$.
\begin{equation}
    H_\text{Er--Gd} = -\frac{2 J_\mathrm{eff}}{\sqrt{N_\mathrm{Gd}/2}} \bm{S}_{\mathrm{Gd}}\cdot\bm{S}.
\end{equation}
The normalisation arises because the erbium only couples to four nearest-neighbour gadolinium ions out of the full $N_\mathrm{Gd}$. The effective exchange constant is seen to be the coupling strength for these four interactions. 

Note that Heisenberg exchange in a crystal-field model is not isotropic if projected down onto isolated spin--½ doublets.

Using the fit that we obtained without gadolinium coupling as a starting point, we fit the high- and low-resolution data to a model with $N_\mathrm{Gd}=2$. The labelling and ordering of transition frequencies was informed by adding coupling into the previous fit and using this to infer by eye which transition belonged to which sublattice. We chose $N_\mathrm{Gd}=2$, because we could clearly identify a small part of a line with two magnon excitations. We used transitions with applied fields up to \qty{1}{T}. Our fitted parameters are given in Table \ref{tab: crystal field}. The other crystal field parameters are not expected to vary much between hosts, so we used the values from Ref.~\cite{carnall_systematic_1989}. Comparing the calculated eigenvalues for the other crystal field levels to measurements from Ref.~\cite{Bertini_2004_S}, we find an root-mean-squared deviation of \qty{12.3}{\per\centi\meter}.

\begin{table}[tb]
\caption{Fitted values of parameters from the crystal field model.}
    \centering
    \begin{tabular}{c r l}
        Parameter & value\hspace{1ex} & \,unit  \\
        \hline\hline
        $E_0$ & $35539.9$ & \unit{\per\centi\meter}  \\
        $F^2$ & $97123.2$ & \unit{\per\centi\meter}  \\
        $F^4$ & $66243.0$ & \unit{\per\centi\meter}  \\
        $F^6$ & $54048.9$ & \unit{\per\centi\meter}  \\
        $\zeta$ & $2362.9$ & \unit{\per\centi\meter}  \\
        $B_2^0$ & $-93.0$ & \unit{\per\centi\meter}  \\
        $B_4^0$ & $316.9$ & \unit{\per\centi\meter}  \\
        $B_6^0$ & $-608.6$ & \unit{\per\centi\meter}  \\
        $B_4^4$ & $768.4$ & \unit{\per\centi\meter}  \\
        $B_6^4$ & $-44.9$ & \unit{\per\centi\meter}  \\
        $E_\mathrm{corr.}$ & $-7.0$ & \unit{\per\centi\meter}  \\
        $B_\mathrm{ex}$ & $-0.28$ & \unit{\tesla}  \\
        $B_\mathrm{dip}$ & $0.32$ & \unit{\tesla}  \\
        $B_\mathrm{0,\mathrm{Gd}}$ & $-1.18$ & \unit{\tesla}  \\
        $J_\mathrm{eff}$ & $-0.38$ & \unit{\per\centi\meter}  \\
        $\theta$ & $6.8$ & degrees  \\
    \end{tabular}
    \label{tab: crystal field}
\end{table}

The relative magnetic-dipole transition strengths were obtained by calculating the magnetic-dipole line strengths
\begin{equation}
S^\mathrm{MD}_{q,f0} = \frac{e^2\hbar^2}{4m^2c^2}|\langle f \rvert L_q + 2S_q \lvert 0 \rangle|^2,
\end{equation}
where $\lvert f\rangle$ is the excited state we are interested in and $\lvert 0 \rangle $ is the ground state.
We note here that we calculated line strengths in a much larger Hilbert space with $N_\mathrm{Gd} = 20$, and a similar correspondence was seen between the observed and calculated magnetic-dipole spectra; the only required change was tweaking $B_{0,\mathrm{Gd}}$ by \qty{15}{\percent}. 

Our fitted misalignment could be an overestimate. Next-nearest-neighbour interactions have been shown to give rich spin structures in other rare-earth antiferromagnets in the same space group \cite{bordelon_frustrated_2021}.

The calculations were carried out using the \texttt{dieke}\cite{dieke}  package in Python.

\renewcommand{\selectlanguage}[1]{}

\end{document}